\documentclass[useAMS,onecolumn,usenatbib]{mn2e}

\newcommand{\prt}{\partial}

\title[Viscosity in spherically symmetric accretion]
{Viscosity in spherically symmetric accretion}
\author[Arnab K. Ray]{Arnab K. Ray\thanks{E-mail:
tpakr@mahendra.iacs.res.in}\\
Department of Theoretical Physics, Indian Association for the
Cultivation of Science, Jadavpur, Calcutta 700032, INDIA}
\begin{document}

%\date{}

%\pagerange{\pageref{}--\pageref{}} \pubyear{}

\maketitle

%\label{firstpage}

\begin{abstract}
The influence of viscosity on the flow behaviour in
spherically symmetric accretion, has been studied here. The
governing equation chosen has been the Navier-Stokes equation. It
has been found that at least for the transonic solution, viscosity
acts as a mechanism that detracts from the effectiveness of
gravity. This has been conjectured to set up a limiting
scale of length for gravity to bring about accretion, and the physical 
interpretation of such a length-scale has been compared with the
conventional understanding of the so-called ``accretion radius" for 
spherically symmetric accretion. For a perturbative presence of viscosity,
it has also been pointed out that the critical points for inflows
and outflows are not identical, which is a consequence of the
fact that under the Navier-Stokes prescription, there is a
breakdown of the invariance of the stationary inflow and outflow
solutions -- an invariance that holds good under inviscid conditions. 
For inflows, the critical point
gets shifted deeper within the gravitational potential well. Finally,
a linear stability analysis of the stationary inflow solutions, 
under the influence of a perturbation that is in the nature of
a standing wave, 
has indicated that the presence of viscosity induces greater
stability in the system, than has been seen for the case of
inviscid spherically symmetric inflows.
\end{abstract}


\section{Introduction}

The influence of viscosity, in a spherically symmetric accreting
system, at least in qualitative terms, has been the object of the study 
that has been undertaken here. Accretion flows are governed by the 
equations of fluid dynamics. This in itself makes a convincing case for 
viscosity to be accorded some importance, since viscosity is an intrinsic
property of fluids. Indeed, 
in accretion processes, viscosity does have a prominent role to play --
as has been established particularly for the axially symmetric case of 
thin accretion discs, where
viscosity, being purportedly a mechanism for the outward transport of
angular momentum, effectively causes the infall of matter. 

However, 
the spherically symmetric accreting system has been studied very widely, 
using only inviscid hydrodynamical equations \citep{bon52, fkr92}. 
In this study, instead of the inviscid Euler equation, the Navier-Stokes 
equation has been considered as one of the governing equations of the flow. 
For a compressible viscous flow with constant coefficients of viscosity,
the Navier-Stokes equation can be rendered in the form \citep{ll87}, 

\begin{equation}
\label{e1}
\frac{\prt {\bf{v}}}{\prt t} +\left({\bf{v}}{\bf{\cdot}}{\bf{\nabla}}\right)
{\bf{v}}
+ \frac{{\bf{\nabla}} P}{\rho} + \frac{GM}{r^2} \hat{\bf r}= \frac{1}{\rho}
\left[\eta {\nabla}^2 {\bf{v}}+ \left(\frac{\eta}{3} + \zeta \right)
{\bf{\nabla}}\left({\bf{\nabla}}{\bf{\cdot}}{\bf{v}}\right)\right]
\end{equation}

For molecular viscosity, for which the actual magnitude may be quite small,
the coefficients of viscosity $\eta$ and $\zeta$, as has been mentioned 
before, have been taken to
be positive constants \citep{ll87}. This greatly simplifies the mathematics,
without largely compromising the underlying physics. In so far as
viscosity is dependent on temperature (which can be connected to a
dependence on the density of the accreting matter), its spatial variation 
is much less compared with that of the velocity of the flow, and in
an actual accreting system where radiative processes facilitate efficient
cooling and keep the system somewhat close to being isothermal, 
it is an expectation that the assumption of constancy of 
$\eta$ and $\zeta$ largely holds true \citep{bh98}.

For the present purpose, $\eta$ and $\zeta$ have been taken to be the
conventional first and second coefficients of viscosity. They are
expected to be of the same order of magnitude \citep{ll87}. Even as
molecular viscosity is understood to be of not much significance in
quantitative terms, its introduction in one of the governing equations
should at least reveal in what qualitative manner would it influence
the flow dynamics -- especially the behaviour of the critical points
of the flows and the stability of the steady inflow solutions under the
effect of a linearized perturbation. 

Considering now the vector identity,

\begin{equation}
\label{e2}
{\bf{\nabla}}{\bf{\times}}\left({\bf{\nabla}}{\bf{\times}}{\bf{v}}\right)
={\bf{\nabla}}\left({\bf{\nabla}}{\bf{\cdot}}{\bf{v}}\right) 
- {\nabla}^2{\bf{v}}
\end{equation}
it is seen that for an irrotational flow, such as in a spherically 
symmetric accreting system, the left-hand side of (\ref{e2}) vanishes and 
leaves behind the condition,

\begin{equation}
\label{e3}
{\bf{\nabla}}\left({\bf{\nabla}}{\bf{\cdot}}{\bf{v}}\right)
={\nabla}^2 \bf{v}
\end{equation}
a result whose use could be made in (\ref{e1}).

It is to be noted here that for a compressible fluid, the equation
of continuity

\begin{equation}
\label{e4}
\frac{\prt \rho}{\prt t} + {\bf{\nabla}}{\bf{\cdot}}
\left(\rho {\bf{v}}\right)=0
\end{equation}
which is the other governing equation of the flow, does not give
the result ${\bf{\nabla}}{\bf{\cdot}}{\bf{v}}=0$, as a steady-state solution.
Comparing this fact with the steady solution of (\ref{e1}), and by using the
identity given in (\ref{e3}), it can be seen that the compressibility of 
accreting matter allows for viscosity to have a role to play, in a 
spherically symmetric flow under steady-state conditions.

\section{The equations of the flow}

For a spherically symmetric system, in which $v\equiv v(r,t)$ and
$\rho \equiv \rho (r,t)$, the equations given 
by (\ref{e1}) and (\ref{e4}) are rendered respectively as,

\begin{equation}
\label{e5}
{\frac{\prt v}{\prt t}}+v{\frac{\prt v}{\prt r}}
+{\frac{1}{\rho}}{\frac{\prt P}{\prt r}}+{\frac{GM}{r^2}}
= \frac{1}{\rho} \left(\frac{4}{3} \eta + \zeta \right)
\frac{\prt}{\prt r}
\left[\frac{1}{r^2} \frac{\prt}{\prt r} \left(vr^2 \right) \right]
\end{equation}
and

\begin{equation}
\label{e6}
{\frac{\prt \rho}{\prt t}}+{\frac{1}{r^2}}{\frac{\prt}{\prt r}}
\left({\rho}vr^2\right)=0
\end{equation}

Use has also been made of a general polytropic equation of state

\begin{equation}
\label{e7}
P=k{\rho}^{\gamma}
\end{equation}
where $\gamma$ is the polytropic exponent with an admissible range
given by $1<{\gamma}<5/3$, the restrictions being imposed by the
isothermal limit and the adiabatic limit, respectively.

The speed of sound $c_s$, with which the velocity of the flow is
suitably scaled, is given by

\begin{equation}
\label{e8}
{c_s}^2={\gamma}k{\rho}^{\gamma -1}
\end{equation}

The set of equations (\ref{e5})-(\ref{e8}), completely describes the 
problem. It can be seen that in (\ref{e5}), the two coefficients of 
viscosity add up like simple scalars, which makes it possible to define 
a total viscosity ${\eta}_{\rm tot}=(4/3)\eta + \zeta $.

In the steady state, explicit time dependence vanishes, 
i.e. $\left({\prt}/{\prt t}\right){\equiv}0$, and from (\ref{e5}) 
and (\ref{e6}) will then follow the results 

\begin{equation}
\label{e9}
v{\frac{dv}{dr}}+{\frac{1}{\rho}}{\frac{dP}{dr}}+{\frac{GM}{r^2}}
={\frac{{\eta}_{\rm tot}}{\rho}}{\frac{d}{dr}}\left[{\frac{1}{r^2}}
{\frac{d}{dr}}\left(vr^2 \right)\right]
\end{equation}
and

\begin{equation}
\label{e10}
{\frac{d}{dr}}\left({\rho}vr^2\right)=0
\end{equation}
respectively. From (\ref{e10}) it further follows that

\begin{equation}
\label{e11}
\frac{1}{r^2}{\frac{d}{dr}}\left(vr^2\right)
=-{\frac{v}{\rho}}{\frac{d \rho}{dr}}
\end{equation}

Using (\ref{e7}), (\ref{e8}) and (\ref{e11}), yields 
from (\ref{e9}) the expression,

\begin{equation}
\label{e12}
\frac{1}{2} \frac{d\left(v^2\right)}{dr} + \frac{1}{\gamma -1} 
\frac{d\left({c_s}^2\right)}{dr}
+f_{\rm eff}=0
\end{equation}
where the effective force $f_{\rm eff}$ is given by,

\begin{equation}
\label{e13}
f_{\rm eff}= \frac{GM}{r^2}- \frac{2}{\gamma -1} 
\frac{{\eta}_{\rm tot}}{\rho}
\frac{d}{dr}\left(-{\frac{v}{c_s}}{\frac{dc_s}{dr}}\right)
\end{equation}

The second expression on the right-hand side of (\ref{e13}) is now 
to be studied closely. For accretion, it is a convention that $v<0$. 
When $r$ decreases, $dc_s/dr$ increases with a negative sign and for 
the transonic solution $\left(-v/c_s \right)$ increases with a 
positive sign. Hence, at least for the transonic solution, the product
$\left(-v/c_s\right)\left(dc_s/dr\right)$ increases with a negative 
sign for decreasing $r$, i.e.

\begin{equation}
\label{e14}
\frac{d}{dr}\left(-{\frac{v}{c_s}}{\frac{dc_s}{dr}}\right)>0
\end{equation}

On using this result in (\ref{e13}), it may be concluded that
$f_{\rm eff}<f_{\rm grav}{\equiv}GM/r^2$. This implies that the 
presence of viscosity detracts from the effectiveness of gravity in 
spherically symmetric accretion. Since it is gravity that drives the 
process of accretion in spherical symmetry, viscosity has an inhibiting 
effect on it, which is quite contrary to what is seen in axially symmetric 
accretion, where viscosity, in effecting the outward transport of angular 
momentum, actually aids the process of accretion \citep{pri81, fkr92}.

It can also be seen that with the introduction of viscosity in a governing
hydrodynamical equation, it would be possible to obtain a limiting scale of
length for the effectiveness of gravity to bring about the infall of matter.
Viscosity is a microscopic phenomenon. Very far away from the accretor,
the bulk influence of gravity, having been greatly weakened, becomes
comparable with the inhibiting influence of viscosity
at the microscopic level, and at such a distance, a scale of length may be
obtained which, in a manner of speaking, may be understood to be a ``viscous 
shielding radius", $r_{\rm visc}$. By dimensional arguments it is also 
possible to have some quantitative understanding of $r_{\rm visc}$. The 
physical quantity ${\dot m}/{{\eta}_{\rm tot}}$ has the dimension of length 
and this quantity may be identified with $r_{\rm visc}$. Here 
${\dot m}$ is the mass accretion rate, which is dependent on the mass of 
the accretor and the ``ambient conditions" \citep{fkr92}, although for 
the viscous system being discussed here, this dependence -- as opposed to
that of the inviscid case, where the dependence is exactly 
known \citep{fkr92} -- may be known only in an order-of-magnitude sense. 
With the identification that $r_{\rm visc}\equiv {\dot m}/{{\eta}_{\rm tot}}$, 
the ``viscous shielding radius" would then conform to what can be intuited 
concerning the role of viscosity, that for a non-zero magnitude of 
${\eta}_{\rm tot}$, the length $r_{\rm visc}$ would set a finite spatial
limit to the accretion process. This understanding of the physical nature of 
$r_{\rm visc}$ can be compared with the somewhat loosely defined concept of 
what has conventionally been known to be the ``accretion radius" \citep{fkr92},
given by $r_{\rm acc}{\cong}GM/{c_s}^2\left(\infty\right)$. 
In all realistic cases with
molecular viscosity, the condition $r_{\rm visc} \gg r_{\rm acc}$ 
prevails. The accreting system 
would then very likely be bound by the scale of length $r_{\rm acc}$. In such
cases, the mass accretion rate $\dot m$, would be very close to the value 
that would be obtained for the inviscid case.  

Viscosity is also expected to influence the critical points of the flow
solutions in the $v^2$ -- $r$ space. 
Critical points of the flow are given by the 
condition that in such a space, for a first-order differential equation
in $v$ and $r$,  
both the numerator and the denominator of 
$d\left(v^2\right)/dr$ vanish simultaneously \citep{skc90}. 
The flow equation (\ref{e9}), however, is a second-order
differential equation. So to get the critical point condition, ideally
a first integral of (\ref{e9}) is to be obtained, which, unfortunately, 
cannot be
done for compressible flows. Hence an approximation is resorted to, that
viscosity has a very small perturbative influence on the conditions
governed by the inviscid equations. For molecular viscosity the validity of
such an approximation is quite evident. Moreover, for large distances, 
the second
derivative of $v$ is comparatively small. The expectation here is that
the critical point would be at a distance where this condition would
hold. A cautionary note to be made here is that for the branch of the
transonic solution very close to the accretor, the second derivative of
$v$ would be large.

Using (\ref{e7}), (\ref{e8}) and (\ref{e10}), would then deliver 
from (\ref{e9}), the result

\begin{equation}
\label{e15}
\left(v-{\frac{{c_s}^2}{v}}-2{\frac{{\eta}_{\rm tot}}{\rho r}}\right)
{\frac{dv}{dr}}=
{\frac{2{c_s}^2}{r}}-{\frac{GM}{r^2}}+{\frac{{\eta}_{\rm tot}}{\rho}}
\left({\frac{d^2 v}{dr^2}}-{\frac{2v}{r^2}}\right)
\end{equation}

The relation ${\rho}={\rho}_{\infty}\left[{c_s}/{c_s}\left(\infty\right)
\right]^{2n}$ is now to be used, where $n=\left(\gamma -1\right)^{-1}$. 
It is then possible to obtain from the right hand side of (\ref{e15}), 
the condition  

\begin{equation}
\label{e16}
{\frac{2{c_{sc}}^2}{r_c}}-{\frac{GM}{{r_c}^2}}+{\frac{{\eta}_{\rm tot}}
{{\rho}_{\infty}}}\left[{\frac{{c_s}\left(\infty\right)}{c_{sc}}}\right]
^{2n}\left({\frac{d^2 v}{dr^2}}{\Big{\vert}}_c - 
{\frac{2v_c}{{r_c}^2}}\right)=0
\end{equation}
where the subscript label $c$, indicates critical point values.
From (\ref{e16}), it is possible to have some idea of the influence of 
viscosity on the critical point of the transonic inflow solution.

A rearrangement of terms in (\ref{e16}) gives,

\begin{equation}
\label{e17}
r_c={\frac{GM}{2{c_{sc}}^2}}+{\frac{{\eta}_{\rm tot}}{{\rho}_{\infty}}}
\left[\frac{c_s\left(\infty\right)}{c_{sc}}\right]^{2n}
{\frac{v_c}{{c_{sc}}^2}}\left(1- \frac{1}{2}
{\frac{{r_c}^2}{v_c}}{\frac{d^2 v}{dr^2}}{\Big{\vert}}_c\right)
\end{equation}

For the transonic solution it may be said that a power law of the
form $v{\sim}r^{- \delta}$ is followed. 
For free fall, $\delta = 1/2$. For the hydrodynamical case being discussed
here, it may be supposed that in the vicinity of the critical point it will
be $1/2<{\delta}<1$. That will make the second term in parentheses in 
(\ref{e17}), a dimensionless positive number less than unity. Since for 
inflows, convention has it that ${v_c}<0$, it can then be said that the 
presence of viscosity causes the critical point to be shifted inwards along 
the $r$-axis, as compared with the inviscid case, 
where ${r_c}=GM/2{c_{sc}}^2$ \citep{fkr92}. This, in itself, is 
consistent with the fact that viscosity 
has the effect of weakening the influence of gravity. At the critical
point, the pressure term balances the gravity term. With gravity having
been weakened, the pressure term is capable of balancing the gravity term 
deeper within the potential well.

The use of the Navier-Stokes equation gives rise to another feature 
of the flow. A look at (\ref{e1}) or its spherically symmetric 
adaptation (\ref{e5}), makes it evident that unlike in the inviscid 
Euler equation, 
$v$ makes an appearance here in the first power (in the viscous term). 
A direct consequence of this fact is that in the steady state, the 
Navier-Stokes equation (\ref{e9}), governing the spherically symmetric
flow, is not invariant under the transformation $v{\longrightarrow}-v$.
The occurrence of $v$ in the first power in (\ref{e9}), causes the 
invariance to break down. The important conclusion to be drawn here is
that unlike in the case of inviscid flows, the critical points of the
transonic inflow and outflow solutions, are not identical in 
the $v^2$ -- $r$
space. It is straightforward to see that in the inviscid limit of
(\ref{e9}), the invariance will be restored \citep{arc99} and the 
conventional critical point conditions for the inviscid case
will be obtained. 

\section{Linear stability analysis of stationary solutions}

The steady state solutions of (\ref{e5}) and (\ref{e6}) 
are $v_0$ and ${\rho}_0$,
on which are imposed small first-order perturbations $v^{\prime}$
and ${\rho}^{\prime}$, respectively. Following a similar prescription
by \cite{pso80} for inviscid spherically symmetric 
accretion, it is convenient
to introduce a new variable, $f{\equiv}{\rho}vr^2$, which in physical 
terms is the mass flux. Its steady-state value $f_0$, as can be seen 
from (\ref{e6}) and (\ref{e10}), is a constant that is identified as 
the mass
accretion rate. A linearized perturbation $f^{\prime}$, about $f_0$ is
given by,

\begin{equation}
\label{e18}
f^{\prime}=r^2\left(v^{\prime}{{\rho}_0}+{v_0}{\rho}^{\prime}\right)
\end{equation}

Linearizing in the perturbation variables $v^{\prime}$ 
and ${\rho}^{\prime}$, gives from (\ref{e6}), the result,

\begin{equation}
\label{e19}
{\frac{\prt {{\rho}^{\prime}}}{\prt t}}+{\frac{1}{r^2}}
{\frac{\prt f^{\prime}}{\prt r}}=0
\end{equation}

With the use of (\ref{e19}), successive differentiation 
of (\ref{e18}) with respect to time yields, first

\begin{equation}
\label{e20}
{\frac{\prt v^{\prime}}{\prt t}}={\frac{1}{{\rho}_0 r^2}}
\left({\frac{\prt f^{\prime}}{\prt t}}
+{v_0}{\frac{\prt f^{\prime}}{\prt r}}\right)
\end{equation}
and then

\begin{equation}
\label{e21}
{\frac{{\prt}^2 v^{\prime}}{\prt t^2}}={\frac{1}{{\rho}_0 r^2}}
\left[{\frac{{\prt}^2 f^{\prime}}{\prt t^2}}+{v_0}{\frac{\prt}{\prt r}}
\left({\frac{\prt f^{\prime}}{\prt t}}\right)\right]
\end{equation}

Now linearizing in terms of the perturbation variables, gives 
from (\ref{e5}) the result

\begin{equation}
\label{e22}
{\frac{\prt v^{\prime}}{\prt t}}+{\frac{\prt}{\prt r}}
\left({v_0}v^{\prime}+{c_{s0}}^2{\frac{{\rho}^{\prime}}{{\rho}_0}}\right)
={\frac{{\eta}_{\rm tot}}
{{\rho}_0}}\left\{{\frac{\prt}{\prt r}}\left[{\frac{1}{r^2}}
{\frac{\prt}{\prt r}}\left(r^2 v^{\prime}\right)\right]
-{\frac{{\rho}^{\prime}}{{\rho}_0}}{\frac{\prt}{\prt r}}
\left[{\frac{1}{r^2}}{\frac{\prt}{\prt r}}\left(r^2 v_0\right)\right]
\right\}
\end{equation}
where $c_{s0}$ is the speed of sound in the steady state.

Partially differentiating (\ref{e22}) with respect to time, and using 
the conditions given by (\ref{e19}), (\ref{e20}) and (\ref{e21}) will 
finally deliver the perturbation equation as

\begin{eqnarray}
\label{e23}
{\frac{{\prt}^2 f^{\prime}}{\prt t^2}}+2{\frac{\prt}{\prt r}}\left({v_0}
{\frac{\prt f^{\prime}}{\prt t}}\right)+{\frac{1}{v_0}}
{\frac{\prt}{\prt r}}\left[{v_0}\left({v_0}^2-
{c_{s0}}^2\right){\frac{\prt f^{\prime}}{\prt r}}\right]
&=& {\eta}_{\rm tot}r^2
\bigg\{{\frac{\prt}{\prt r}}\left[{\frac{1}{r^2}}{\frac{\prt}{\prt r}}
\left({\frac{1}{{\rho}_0}}{\frac{\prt f^{\prime}}{\prt t}}+{\frac{v_0}
{{\rho}_0}}{\frac{\prt f^{\prime}}{\prt r}}\right)\right] \nonumber \\
& & + {\frac{1}{{\rho}_0 r^2}}{\frac{\prt f^{\prime}}{\prt r}}{\frac{\prt}
{\prt r}}\bigg[{\frac{1}{r^2}}{\frac{\prt}
{\prt r}}\left(r^2 v_0\right)\bigg]\bigg\}
\end{eqnarray}

It is quite evident that the related perturbation equation for the
inviscid case \citep{td92}, will be readily obtained for 
${\eta}_{\rm tot}=0$.

For (\ref{e23}), a solution of the form
$f^{\prime}=g(r)e^{\Omega t}$ is chosen,
in which $g(r)$ is the spatial part of the perturbation and $\Omega$
in general is complex. This leads to a quadratic equation 
in $\Omega$, which is 

\begin{equation}
\label{e24}
{\rm A}{\Omega}^2+{\rm B}{\Omega}+{\rm C}=0
\end{equation}
where

\begin{eqnarray}
\label{e25}
{\rm A} &=& g \nonumber \\
{\rm B} &=& 2{\frac{d}{dr}}\left(v_0 g \right) - {\eta}_{\rm tot}r^2
{\frac{d}{dr}}\left[{\frac{1}{r^2}}{\frac{d}{dr}}\left(
{\frac{g}{{\rho}_0}}\right)\right] \nonumber \\
{\rm C} &=& {\frac{1}{v_0}}{\frac{d}{dr}}\bigg[v_0\left({v_0}^2-
{c_{s0}}^2\right){\frac{dg}{dr}}\bigg]
-{\eta}_{\rm tot} \left\{
r^2 \frac{d}{dr} \left[ \frac{1}{r^2} \frac{d}{dr}
\left( \frac{v_0}{{\rho}_0} \frac{dg}{dr} \right)\right]+
\frac{1}{{\rho}_0} \frac{dg}{dr} \frac{d}{dr} \bigg[
\frac{1}{r^2} \frac{d}{dr} \left( r^2 v_0\right)\bigg]\right\}
\end{eqnarray}

Stability of the subsonic inflows has been studied here with a particular
emphasis. In such flows, it is readily understandable that for a 
perturbation which is in the form of a standing wave, there are
two spatial points, one very close to the accretor and one very far
away from it, where it is possible to have boundary conditions which
would constrain the perturbation to die out \citep{pso80}. 
Multiplying (\ref{e24})
by $v_0 g$ and integrating over the range of $r$, that is bounded by
the two points where the perturbation dies out, will give,

\begin{equation}
\label{e26}
{\cal{A}}{\Omega}^2 + {\cal{B}}{\Omega} + {\cal{C}} = 0
\end{equation}
where

\begin{eqnarray}
\label{e27}
{\cal{A}} &=& \int v_0 g^2 \, dr \nonumber \\
{\cal{B}} &=& {\frac{4 \pi {\eta}_{\rm tot}}{\left(-{\dot{m}}\right)}} 
\int r^{-2}\bigg[{\frac{d}{dr}}\left(v_0 gr^2\right)\bigg]^2 \, dr 
\nonumber \\
{\cal{C}} &=& - \int v_0\left({v_0}^2-{c_{s0}}^2\right)
\left( \frac{dg}{dr}\right)^2 \, dr 
-{\eta}_{\rm tot} \int \frac{v_0 g}{{\rho}_0}
\left(\frac{dg}{dr}\right) \frac{d}{dr}\left(- \frac{v_0}{{\rho}_0} 
\frac{d{\rho}_0}{dr} \right) \, dr
-{\eta}_{\rm tot} \int r^{-2} \frac{d}{dr} \left(v_0 gr^2 \right) 
\frac{d}{dr} \left(- \frac{v_0}{{\rho}_0} \frac{dg}{dr} \right) \, dr
\end{eqnarray}

The results given by (\ref{e26}) and (\ref{e27}) have been arrived at 
by introducing the simplification of partial integrations, and by 
requiring that all terms at the two boundaries vanish \citep{pso80}. 
Use has also been made of an integrated form of (\ref{e10}), which 
is ${\rho}_0 v_0 r^2 = \left(-{\dot m}\right)/4 \pi$,
for spherically symmetric inflows \citep{fkr92}. The 
integration constant ${\dot m}$ is physically identified with the 
mass infall rate and therefore ${\dot m}>0$.

The solution of (\ref{e26}) is given by,

\begin{equation}
\label{e28}
\Omega= -\frac{\cal B}{2\cal A} \pm \sqrt{{\frac{{\cal{B}}^2}
{4{\cal{A}}^2}}-\frac{\cal C}{\cal A}}
\end{equation}

The result in (\ref{e28}) indicates that $\Omega$ has a real part and to
ensure stability it is necessary that $\Re\left({\Omega}\right)<0$. An 
examination of $\cal A$ and $\cal B$ will give,

\begin{equation}
\label{e29}
\frac{\cal B}{\cal A} = \frac{4 \pi {\eta}_{\rm tot}}
{\left(-{\dot m}\right)}\left({\int} v_0 g^2 \, dr \right)^{-1} 
\int r^{-2}\bigg[{\frac{d}{dr}}\left(v_0 gr^2 \right)\bigg]^2 \, dr 
\equiv {\frac{4 \pi {\eta}_{\rm tot}}{{\dot{m}}}} 
\left[\int \left(-v_0 \right)g^2 \, dr \right]^{-1} \int r^{-2}
\bigg[{\frac{d}{dr}}\left(v_0 gr^2 \right)\bigg]^2 \, dr 
\end{equation}

For inflows, the condition is $\left(-v_0 \right)>0$, which, for a 
real $g$, would then imply that $\left({\cal B}/{\cal A}\right)>0$. 
Instabilities could still arise if $\left(-{\cal C}/{\cal A}\right)>0$, 
because $\Omega$ would then have one positive root. It is seen that

\begin{eqnarray}
\label{e30}
-\frac{\cal C}{\cal A} &=& \bigg[ \int v_0 \left({v_0}^2-{c_{s0}}^2 \right)
\left(\frac{dg}{dr}\right)^2 \, dr
+{\eta}_{\rm tot} \int \frac{v_0 g}{{\rho}_0}\left(\frac{dg}{dr}\right)
\frac{d}{dr}\left(- \frac{v_0}{{\rho}_0}{\frac{d{\rho}_0}{dr}}\right)\,dr
\nonumber \\
& & +{\eta}_{\rm tot} \int r^{-2} \frac{d}{dr} \left(v_0 gr^2 \right) 
\frac{d}{dr}\left(- \frac{v_0}{{\rho}_0}{\frac{dg}{dr}}\right)\,dr
\bigg] \left(\int v_0 g^2\, dr \right)^{-1} 
\end{eqnarray}

In the inviscid term of (\ref{e30}), for subsonic inflows, the governing
condition is ${v_0}^2<{c_{s0}}^2$. 
This would make the term always negative for any real $g$. For the 
viscous terms, the asymptotic properties of the integrands near the 
outer boundary, would have to be studied. For $r{\longrightarrow}{\infty}$,
the boundary conditions are ${\rho}_0{\longrightarrow}{\rho}_{\infty}$, 
$v_0{\longrightarrow}0$ and $g{\longrightarrow}0$. In that case, for the 
first viscous term, asymptotically at least, the products of the 
derivatives within the integrand, would be negative. For the second 
viscous term, the integrand in the numerator would be positive 
asymptotically, but having a negative denominator (since $v_0<0$), the 
term would be negative overall. It is to be expected that the conclusion
drawn on the basis of the asymptotic argument, could be logically
extended over the entire range of the integration. This would then lead
to the conclusion that $\left(-{\cal C}/{\cal A}\right)<0$ for all 
situations of physical interest, and thus makes it possible to 
avoid having instabilities to develop in the system. 

It is to be noted here that for inviscid subsonic inflows, a similar
analysis would give the result that $\Omega$ would be purely imaginary
\citep{pso80}, i.e. the perturbation would be oscillatory with a constant
amplitude. The stability analysis for the viscous flow solutions developed 
above, indicates that even if the perturbation were to be oscillatory, its
amplitude would decrease. Viscosity is fluid friction in its essence and 
as such it damps out the amplitude of the perturbation. 
This would therefore establish that the presence of viscosity would 
induce greater stability in the system. 
This is a result that is somewhat surprising, in view of the fact that 
viscosity is a 
dissipative mechanism that releases heat into the system. To understand
this seeming anomaly, it would be worthwhile to go back to (\ref{e9}), in 
which, the viscous term gives a mode for the loss of mechanical energy from 
the system. However, in the choice of equations here, no account has been 
taken of an equation for the energy balance, that, for viscous dissipation 
of the mechanical energy, would monitor the concomitant rise
in the internal energy of the system. This loss of mechanical energy 
manifests itself as greater stability for the system, compared with 
the inviscid case. 

For the stability analysis of the transonic solution, a different line
of reasoning would have to be adopted, because there is no physically
feasible inner boundary condition for the inflow 
\citep{pso80}, and neither is the condition ${v_0}^2<{c_{s0}}^2$ satisfied 
within the sonic radius. 
\citet{td92} have shown that in the case of inviscid flows, for the subsonic
branch of the transonic solution, the flow stability is dependent on
conditions prevailing at the outer boundary. For the supersonic branch,
recourse is to be had to the argument of \citet{gar79} that for a finite 
supersonic region, a disturbance would be carried away in a finite time, and 
in this manner would ensure the stability of the solution. These arguments, 
it is to be expected, would hold good for the transonic solution for viscous
flows. The expectation is based on the analogous reasoning that for
the inviscid flow solutions, the conclusions drawn about the stability 
of the subsonic inflows are largely and qualitatively the same as those
about the stability of the transonic inflow. More so, when, for viscous 
flow solutions, it has been argued that the system shows itself to have 
greater stability for the subsonic stationary
inflow solutions under the influence of a linearized perturbation. Hence,
it may be safely established that a linear stability analysis 
indicates that in the absence of any energy balance equation, viscosity,
as incorporated in the Navier-Stokes equation, would apparently induce 
greater stability in the stationary solutions in a spherically symmetric 
accreting system. 

\section{Concluding remarks}

It has been seen that the presence of viscosity in a spherically 
symmetric accreting system, adversely affects the effectiveness of 
gravitation in bringing about the infall of matter -- at least for the
transonic solution. To the extent that the transonic solution is the
one that is most likely to be realizable \citep{bon52, gar79, rb02}, 
the effect
of viscosity on such flows can have important implications -- especially
so, when a significantly large and scale-dependent effective 
``turbulent viscosity" is taken into consideration.
Such a prescription is not entirely without its context or relevance, 
because spherically symmetric infall of matter is expected to be seen
in accretion from the interstellar medium on to an isolated accretor,
and the interstellar medium, as has been well known, displays turbulent
behaviour. And since the Navier-Stokes equation is closely connected to
turbulence \citep{fri99}, the extension to the use of an 
effective turbulent viscosity to study a spherically symmetric accreting
system, acquires some justification for itself.

Viscosity, as has been discussed in an earlier section, could possibly set 
up a limiting length-scale for gravity to be effective enough in driving 
a spherically symmetric accreting system. Yet it can be seen that for a 
weakly viscous system, the ``viscous shielding radius" $r_{\rm visc}(\equiv 
{\dot m}/{{\eta}_{\rm tot}})$ would scarcely be able to set a reasonable 
spatial bound on the accretion process. In such a situation the accreting
system would be spatially limited by the accretion radius $r_{\rm acc}$. 
However, the theoretical possibility is open that
$r_{\rm visc}$ could become comparable with $r_{\rm acc}$ with the 
introduction of a significantly large and scale-dependent turbulent viscosity, 
which could then make the viscous shielding radius define a noticeable limit 
on the active spatial range of the accretion process. The relevance of 
prescribing a scale-dependent turbulent viscosity becomes quite 
evident, when it is to be recalled once again that the interstellar medium, 
which lends itself readily to spherical accretion, is widely acknowledged
to be a turbulent system. 

It has been seen here that if viscosity is considered to have a small
perturbative influence on the conditions governed by inviscid equations,
then the critical point for the inflow branch is marginally shifted
deeper within the potential well. It is a matter worth some serious 
conjecture, as to how far reaching the consequences would be for the
critical point, if instead of a small perturbative viscosity, a significantly 
large effective turbulent viscosity were to be imposed. Since, the
transonic inflow solution passes through the critical point, its 
behaviour could get radically affected. 

Much the same thing can be surmised for the result of a linear
stability analysis of the steady flow solutions, with a turbulent
viscosity prescribed for the Navier-Stokes equation. The change could be
even more significant, considering that such an effective turbulent viscosity, 
being scale-dependent, would have a spatial variation 
\citep{ms77}
-- not to mention factoring in the accompanying rise in the internal 
energy of the system. 

\section*{Acknowledgments}

This research has made use of NASA's Astrophysics Data System.
The financial assistance provided by the Council of Scientific and
Industrial Research, Government of India, is being gratefully 
acknowledged here. Gratitude is also to be expressed to Prof. J.K.
Bhattacharjee for his very helpful advice and comments. The suggestion
of an anonymous referee that the term ``viscous shielding radius" 
should be introduced is also being acknowledged.

\bsp

\label{lastpage}

\end{document}